\documentclass[preprint,showpacs,preprintnumbers,amsmath,amssymb,endfloats]{revtex4}
\usepackage{graphicx}
\usepackage{dcolumn}
\usepackage{bm}
\usepackage{amssymb}
\usepackage{amsmath}
\usepackage{latexsym}
\usepackage{longtable}
\usepackage{ulem}

\begin{document}

\title{Why does neutron transfer play  different roles  in  sub-barrier fusion reactions $^{32}$S+$^{94,96}$Zr and $^{40}$Ca+$^{94,96}$Zr?}
\author{V.V.Sargsyan$^{1,2}$, G.G.Adamian$^{1}$, N.V.Antonenko$^1$, W. Scheid$^3$, and  H.Q.Zhang$^4$
}
\affiliation{$^{1}$Joint Institute for Nuclear Research, 141980 Dubna, Russia\\
$^{2}$International Center for Advanced Studies, Yerevan State University, M. Manougian 1, 0025, Yerevan, Armenia\\
$^{3}$Institut f\"ur Theoretische Physik der Justus--Liebig--Universit\"at, D--35392 Giessen, Germany\\
$^{4}$China Institute of Atomic Energy, Post Office Box 275, Beijing 102413,  China
}
\date{\today}

\begin{abstract}
The sub-barrier capture (fusion) reactions
$^{32}$S+$^{90,94,96}$Zr, $^{36}$S+$^{90,96}$Zr, $^{40}$Ca+$^{90,94,96}$Zr,  and  $^{48}$Ca+$^{90,96}$Zr
with   positive and negative $Q$-values for  neutron transfer are
studied   within the quantum diffusion approach and the universal fusion function representation. For these systems,
the $s$-wave capture probabilities are extracted from the experimental excitation functions  and are also analyzed.
Different effects of the positive $Q_{xn}$-value neutron transfer in the fusion   enhancement  are revealed
in the relatively close reactions $^{32}$S+$^{94,96}$Zr and $^{40}$Ca+$^{94,96}$Zr.

\end{abstract}


 \maketitle

\section{Introduction}
The  nuclear deformation effects are identified as playing a major role in the
magnitude of the sub-barrier fusion (capture)  cross sections \cite{Gomes,Back}.
There are a several experimental evidences which confirm the straightforward influence
of  nuclear deformation on the  fusion.
If the target nucleus is prolate in the ground state, the Coulomb field on its tips is lower than on its sides. Thus,
 the capture or fusion probability  increases at  energies below the barrier corresponding to the spherical nuclei.

The dynamics of neutron  transfer-mediated sub-barrier capture and fusion is not yet revealed~\cite{Back}.
The cross section enhancement in the sub-barrier fusion
of $^{58}$Ni+$^{64}$Ni, with respect to $^{58}$Ni+$^{58}$Ni,~\cite{Beckerman}
is interpreted in Ref. \cite{Broglia} as a kinematic effect
due to the positive $Q_{2n}$-value of the ground-state-to-ground-state two-neutron transfer ($2n$-transfer)
channel.
A correlation is observed between  considerable  sub-barrier fusion  enhancement and positive $Q_{xn}$-values for neutron transfer in
the reactions
$^{40}$Ca+$^{94,96}$Zr~\cite{TimmersCa40Zr96,StefaniniCa40Zn94,StefMont40ca96zr}, and
$^{40}$Ca+$^{116,124,132}$Sn~\cite{Stefanini40ca116124sn,Kol40Ca124132Sn}.
%
%
%
%

The importance of neutron transfer with positive $Q_{xn}$-values in nuclear fusion (capture) originates from
the fact that neutrons are insensitive to the Coulomb barrier and
their transfer   starts at quite larger separations,
before the projectile is captured by target-nucleus.
It is generally thought that  the sub-barrier capture
(fusion) cross section increases because of the neutron
transfer~\cite{TimmersCa40Zr96,StefaniniCa40Zn94,StefMont40ca96zr,Stefanini40ca116124sn,Kol40Ca124132Sn,Pengo,Henning,Stelson,
Roberts,Stefanini3236s110pd,Sonzogni,Tripathi,Jiang40ca48ca,ZhangS32Zr9096,Kola40ca134Te,Jia32s94zr}.
However, the reduced excitation functions for the
reactions $^{16,17,18}$O+$^{A}$Sn ($A$=112,116-120,122,124)~\cite{AOASn},
scaled to remove the effects of smoothly varying barrier parameters, do not show any
strong dependence on the mass number of target or projectile.
The relative changes are within a factor two and are not correlated
with the positive $Q_{xn}$-values of neutron-transfer channels in these reactions.
As   shown   in Ref.~\cite{Scarlassara},
the neutron transfer channels with  positive $Q_{xn}$-value
weakly influence  the capture (fusion) cross section
in the  $^{60}$Ni+$^{100}$Mo
reaction at sub-barrier energies.
In the reactions $^{40}$Ca+$^{116,124}$Sn ($Q_{2n}>0$) and
$^{132}$Sn,$^{130}$Te+$^{58,64}$Ni ($Q_{2n}>0$)
at energies above and a few MeV below
the Coulomb
barrier, the effect of transfer channels on
the capture (fusion)  is demonstrated to be very weak with no
significant differences observed in the reduced excitation
functions~\cite{Stefanini40ca116124sn,Liang}.
In comparison with the   $^{16}$O+$^{76}$Ge reaction \cite{Jia18O74Ge}, 
the fusion enhancement due to the positive $Q_{2n}$-value
is not   revealed
in the $^{18}$O+$^{74}$Ge reaction.

It is presently not clear why the neutron transfers with positive $Q_{xn}$-values play a decisive role  in the fusion reactions
$^{40}$Ca+$^{48}$Ca, $^{58}$Ni+$^{64}$Ni, $^{40}$Ca+$^{94,96}$Zr, $^{40}$Ca+$^{116,124,132}$Sn
and weakly influence  the fusion reactions $^{58,64}$Ni+$^{132}$Sn, $^{58,64}$Ni+$^{130}$Te, $^{60}$Ni+$^{100}$Mo, $^{18}$O+$^{74}$Ge, $^{18}$O+$^{A}$Sn \cite{Back,Jiang2014}.
Although the enhancement appears to be related to the existence of large positive $Q_{xn}$-values for neutron transfer,
it is not proportional to the magnitudes of those $Q_{xn}$-values,
which are  larger for $^{40}$Ca+$^{96}$Zr [$^{40}$Ca+$^{132}$Sn or $^{40}$Ca+$^{124}$Sn] than for $^{40}$Ca+$^{94}$Zr [$^{40}$Ca+$^{124}$Sn or $^{40}$Ca+$^{116}$Sn]. The sub-barrier enhancements are similar in these reactions.
So, the influence of neutron transfer on the capture process is not trivial to be easily explained.

The quantum diffusion approach~\cite{EPJSub,EPJSub1,EPJSub3,EPJSub21,SD} was applied to study
the role of the neutron transfer with   positive $Q_{xn}$-value
in the capture (fusion) reactions  at sub-, near- and above-barrier energies.
 A good agreement of the theoretical calculations with the  experimental data was demonstrated .
As found, the change of   the capture cross section after the neutron transfer
occurs due to the change of the deformations of  nuclei~\cite{EPJSub,EPJSub1,EPJSub3,EPJSub21,SD}.
Thus, the  effect of the  neutron transfer is an indirect influence of the quadrupole
deformation.
As demonstrated in Ref. \cite{EPJSub1},
the neutron transfer can weakly influence or even suppress the capture (fusion)  cross section in some reactions.

Applying the quantum diffusion approach~\cite{EPJSub,EPJSub1,EPJSub3,EPJSub21,SD} (Sect. IV),
the universal fusion function representation~\cite{Gomes1,Gomes2} (Sect. II), and capture probabilities extracted from the experimental
excitation functions (Sect. III),
 we try to  answer  the question how the  neutron transfer influence  the sub-barrier capture     cross section in the reactions
$^{32}$S+$^{90,94,96}$Zr, $^{36}$S+$^{90,96}$Zr, $^{40}$Ca+$^{90,94,96}$Zr,  and  $^{48}$Ca+$^{90,96}$Zr  at near and sub-barrier energies.
We will show why the influence of positive $Q_{xn}$-value  neutron transfer
is completely different in the relatively close reactions $^{32}$S+$^{94,96}$Zr and $^{40}$Ca+$^{94,96}$Zr.


\section{Experimental reduced capture cross sections}
To analyze the capture cross sections in the reactions with
different Coulomb barrier heights $V_b$ and radius $R_b$
calculated in the case of spherical nuclei,
it is useful to compare not the excitation functions,
but the dependence of the dimensionless quantities
$2E_{\rm c.m.}\sigma_{cap}(E_{\rm c.m.})/(\hbar\omega_bR_{b}^{2})$
versus
$(E_{\rm c.m.}-V_b)/(\hbar\omega_b)$
or
$(E_{\rm c.m.}-V_b)$~\cite{Gomes1,Gomes2}.
Here,
$\omega_b$ and $\mu $ are the frequency of an inverted oscillator approximated the
barrier    
and the reduced mass of the system, respectively.
In the reactions, where the capture and fusion
cross sections coincide, the comparison of experimental data with
the universal fusion function~\cite{Gomes1,Gomes2} allows us to  
conclude about the role of static deformations of the colliding nuclei
and the nucleon transfer between them in the capture cross section. Indeed,
the universal function disregards these effects.

For the reactions
$^{40}$Ca+$^{90}$Zr,
$^{48}$Ca+$^{90,96}$Zr,
 and
$^{36}$S+$^{90,96}$Zr,
with almost spherical nuclei
and without neutron transfer
[the negative $Q_{xn}$-values],
the experimental cross sections are rather close and fall with the same rate like the
universal fusion function (Fig.~1).
For the reactions
$^{40}$Ca+$^{94,96}$Zr
with the  neutron transfer [the positive $Q_{xn}$-values], one can see
that the reduced cross sections  strongly deviate from the
universal function   in contrast to the reactions
$^{40}$Ca+$^{90}$Zr,
$^{36}$S+$^{90,96}$Zr,
$^{48}$Ca+$^{90,96}$Zr,
where the neutron transfer
is suppressed.

In the reactions
$^{32}$S+$^{90,94,96}$Zr
with strongly deformed projectile $^{32}$S (Fig.1),
the deviations from the
universal function are  mainly caused by the static deformation effects.
In spite of the $Q_{xn}$-values for the neutron transfer
range from the negative [$^{32}$S+$^{90}$Zr] to large and positive values [$^{32}$S+$^{94,96}$Zr],  
 the reduced capture (fusion) cross sections appear to be almost the same.
So,  we observe the strong
and weak influence of neutron transfer on the capture cross sections in the reactions 
$^{40}$Ca+$^{94,96}$Zr and $^{32}$S+$^{94,96}$Zr, respectively.

\section{Capture probabilities extracted from experimental capture excitation functions}
Shifting the energy by the rotational energy $E_R(J)=\frac{\hbar^{2}J(J+1)}{2\mu R_{b}^{2}}$ \cite{Bala},
one can approximate the angular momentum $J$ dependence of the transmission (capture) probability $%
P_{cap}(E_{\mathrm{c.m.}},J)$, at a given $E_{\mathrm{c.m.}}$:
\begin{equation}
P_{cap}(E_{\mathrm{c.m.}},J)\approx P_{cap}(E_{\mathrm{c.m.}}-E_R(J),J=0).
\end{equation}%
If we use the formula for the capture cross section,
convert the sum over the partial waves $J$ into an integral, and express $J$
by the variable $E=E_{\mathrm{c.m.}}-E_R(J)$,
 we obtain the following simple expression:
\begin{equation}
\sigma _{cap}(E_{\mathrm{c.m.}})=\frac{\pi R_{b}^{2}}{E_{\mathrm{c.m.}}}%
\int_{0}^{E_{\mathrm{c.m.}}}dEP_{cap}(E,J=0).
\end{equation}
Multiplying this equation by $E_{\mathrm{c.m.}}/(\pi R_{b}^{2})$ and
differentiating over $E_{\mathrm{c.m.}}$, one obtains \cite{Bala}:
\begin{equation}
P_{cap}(E_{\mathrm{c.m.}},J=0)=\frac{1}{\pi R_{b}^{2}}\frac{d[E_{\mathrm{c.m.%
}}\sigma _{cap}(E_{\mathrm{c.m.}})]}{dE_{\mathrm{c.m.}}}.
\label{g3_eq}
\end{equation}
One can see that $\frac{d[E_{\mathrm{c.m.}}\sigma _{cap}(E_{%
\mathrm{c.m.}})]}{dE_{\mathrm{c.m.}}}$ has a meaning  
of the $s$-wave transmission in the entrance channel.
Therefore,
 the $s$-wave capture probability can be extracted with a satisfactory accuracy from
the experimental capture cross sections $\sigma _{cap}(E_{\mathrm{c.m.}})$
at energies near and below the Coulomb barrier.
Note that at energies considered the dependence of the Coulomb barrier radius on the
angular momentum is very weak.

The extraction method just described requires some procedure to smooth the experimental data
since the values of $E_{\mathrm{c.m.}}\sigma _{cap}(E_{\mathrm{c.m.}})$ have error bars.
We spline the experimental points of  $E_{\mathrm{c.m.}}\sigma _{cap}(E_{\mathrm{c.m.}})$ by the B\'ezier parametric curve \cite{Bezier}.

In Figs. 2 and 3,
the extracted capture probabilities $P_{cap}(E,J=0)$ demonstrate
the influence of nucleon transfer on the capture (fusion)
excitation function.
In the  reactions $^{40}$Ca+$^{90}$Zr and
$^{48}$Ca+$^{90,96}$Zr  with the negative $Q_{xn}$-values  for nucleon transfer, the capture probability
exhibits a steep falloff of the probability at low energies. However,
the first derivatives of $P_{cap}(E,J=0)$ are almost the same.
Conversely, the reactions  $^{40}$Ca + $^{94,96}$Zr have positive $Q_{xn}$-values for neutron transfer. This leads to
the smaller slope of probability functions at sub-barrier energies. The capture 
probabilities in the reactions $^{40}$Ca+$^{94,96}$Zr are close to each other.

Since the nucleus $^{36}$S is spherical, the slopes of functions $P_{cap}(E,J=0)$
for the
reactions
$^{36}$S+$^{90,96}$Zr are  larger than those for the reactions $^{32}$S+$^{94,96}$Zr and $^{32}$S+$^{90}$Zr with the strongly deformed $^{32}$S.
The slopes of functions $P_{cap}(E,J=0)$ are rather similar (Fig. 3)
in the $^{32}$S+$^{90}$Zr reaction with the negative $Q_{xn}$-values for
neutron transfer and in the reactions  $^{32}$S+$^{94,96}$Zr with the positive $Q_{xn}$-values for
neutron transfer.
Thus, the enhancement of capture probability in these reactions has the same origin. It arises
due to the large static deformations of nuclei   $^{32,34}$S and the neutron transfer 
is not responsible for the  capture (fusion)
enhancement.

As follows from the extracted capture probabilities, the experimental normalizations of the cross sections are different
in the reactions  $^{32}$S+$^{90,94,96}$Zr  and   $^{36}$S+$^{90,96}$Zr.
One should think about the experimental reasons for such deviations.


\section{Calculations within the quantum diffusion approach}
In the quantum diffusion approach~\cite{EPJSub,EPJSub1,EPJSub3,EPJSub21,SD,PRCPOP}
the collisions of  nuclei are described with
a single relevant collective variable: the relative distance  between
the colliding nuclei. This approach takes into consideration the fluctuation and dissipation effects in
collisions of heavy ions which model the coupling with various channels
(for example, the non-collective single-particle excitations, low-lying collective dynamical modes
 of the target and projectile).
We have to mention that many quantum-mechanical and non-Markovian effects accompanying
the passage through the potential barrier are taken into consideration in our
formalism~\cite{EPJSub,EPJSub1,EPJSub3,EPJSub21,SD,PRCPOP}.
The  nuclear deformation effects
are taken into account through the dependence of the nucleus-nucleus potential
on the deformations and mutual orientations of the colliding nuclei.
To calculate the nucleus-nucleus interaction potential $V(R)$,
we use the procedure presented in Refs.~\cite{EPJSub,EPJSub1,EPJSub3,EPJSub21,SD,PRCPOP}.
For the nuclear part of the nucleus-nucleus
potential, the double-folding formalism with
the Skyrme-type density-dependent effective
nucleon-nucleon interaction is used.
With this approach many heavy-ion capture
reactions at energies above and well below the Coulomb barrier have been
successfully described~\cite{EPJSub,EPJSub1,EPJSub3,EPJSub21,SD,PRCPOP}.

Following the hypothesis of Ref.~\cite{Broglia},
we assume that the sub-barrier capture
in the reactions under consideration
mainly  depends  on the two-neutron
transfer with the   positive   $Q_{2n}$-value.
Our assumption is that, just before the projectile is captured by the target-nucleus
(just before the crossing of the Coulomb barrier) which is a slow process,
the $2n$-transfer ($Q_{2n}>0$)  transfer  occurs   and  leads to the
population of the first excited collective state in the recipient nucleus~\cite{SSzilner}
(the donor nucleus remains in the ground state).
The absolute values of the quadrupole deformation parameters $\beta_2$ in 2$^+$ state
of even-even deformed nuclei are taken from Ref.~\cite{Ram}.
For the  nuclei deformed in the
ground state, the $\beta_2$ in the first excited collective state is similar
to that in the ground state.
For the double magic and semi-magic nuclei,
we take $\beta_2=0$ in the ground state.

The motion to
$N/Z$ equilibrium starts in the system before the capture occurs because it is energetically favorable
in the dinuclear system in the vicinity of the Coulomb barrier.
For the reactions under consideration,
the average change of mass asymmetry is related to the two-neutron
transfer. In these reactions, $Q_{2n}>Q_{1n}$ and during the capture the
$2n$-transfer is more probable than $1n$-transfer.
After the $2n$-transfer the mass numbers,  the deformation parameters
of the interacting nuclei, and, correspondingly, the height $V_b$
and shape of the Coulomb barrier   change. Then
one can expect an enhancement or suppression of the capture.
If  after the neutron transfer the deformations of interacting nuclei increase (decrease),
the capture probability increases (decreases).
If  after the transfer the deformations of interacting nuclei do not change,
there is no effect of the neutron transfer on the capture.
This scenario was verified in the description of many  reactions~\cite{EPJSub,EPJSub1,EPJSub3,EPJSub21,SD,PRCPOP}.

In Fig.~4  one can see a good agreement between the  calculated  and the experimental  capture cross sections
in the reactions $^{40}$Ca+$^{94,96}$Zr  with the positive $Q$ values for
neutron transfer and  in the reactions $^{40}$Ca+$^{90}$Zr, $^{48}$Ca+$^{90,96}$Zr with  negative $Q$-values for
neutron transfer.
%
The theoretical calculations describe the strong  deviation of
the slopes of excitation functions   in the  reactions $^{40}$Ca+$^{94,96}$Zr
with  positive $Q$-values for neutron transfer from   those in the reactions
$^{40}$Ca+$^{90}$Zr, $^{48}$Ca+$^{90,96}$Zr,
where the neutron transfers   are suppressed because of  negative $Q$-values.
This means that the observed capture enhancements  in
the reactions $^{40}$Ca+$^{94,96}$Zr  at sub-barrier energies
are related to the two-neutron transfer effect.
 After $2n$-transfer in the reactions\\
 $^{40}$Ca($\beta_2=0$)+$^{94}$Zr($\beta_2=0.09$)$\to ^{42}$Ca($\beta_2=0.25$)+$^{92}$Zr($\beta_2=0.1$)
[$Q_{2n}$=4.9 MeV]\\
 and\\
$^{40}$Ca($\beta_2=0$)+$^{96}$Zr($\beta_2=0.08$)$\to ^{42}$Ca($\beta_2=0.25$)+$^{94}$Zr($\beta_2=0.09$)
[$Q_{2n}$=5.5 MeV],\\
the deformation of the light nucleus strongly increases   and,
thus, the height of the Coulomb barrier decreases and
the capture cross section becomes larger (Fig.~4). So, because of the neutron-transfer effect
the reactions $^{40}$Ca+$^{94,96}$Zr show large sub-barrier enhancements with respect to
the reactions $^{48}$Ca+$^{90,96}$Zr and $^{40}$Ca+$^{90}$Zr.
One can see in Fig.5 that with decreasing the sub-barrier energy
the cross sections with and without two-neutron transfer   strongly deviate.
The slopes of the excitation functions
in the reactions
$^{40}$Ca+$^{94,96}$Zr are almost the same because in both cases
after the neutron transfer the nuclei have similar deformations.
The relative enhancement of the sub-barrier fusion cross sections in
the reactions
$^{40}$Ca+$^{94,96}$Zr with respect to those in the reactions
$^{48}$Ca+$^{90,96}$Zr and $^{40}$Ca+$^{90}$Zr is mainly related to the deformation
of $^{42}$Ca in the $2^+$ state.
Thus, the observed capture enhancement at  sub-barrier energies in
the reactions $^{40}$Ca+$^{94,96}$Zr  is purely  related to the  transfer effects.

Since the sub-barrier enhancements are surprisingly similar for the two reactions  $^{40}$Ca+$^{94,96}$Zr
with different positive $Q$-values for neutron transfer, one can assume that the absolute value of the positive $Q$-value is  rather unimportant for the capture
following transfer. If the transfer is energetically favorable it occurs during
the capture process.  In this case the transfer influences the capture (fusion)
[or the height and width of the Coulomb barrier]  through the change of the isotopic composition
of interacting nuclei
and, correspondingly, through the change of their deformations.

Figure   6  shows the capture (fusion) excitation functions  for the reactions $^{32}$S+$^{90,94,96}$Zr and $^{36}$S+$^{90,96}$Zr.
The $Q_{2n}$-values for the $2n$-transfer
processes are positive (negative) for the reactions $^{32}$S+$^{94,96}$Zr ($^{32}$S+$^{90}$Zr, $^{36}$S+$^{90,96}$Zr).
After the $2n$-transfer (before the capture)
in the reactions (Figs.~4 and 6)\\
$^{32}$S($\beta_2=0.31$)+$^{94}$Zr($\beta_2=0.09$)$\to ^{34}$S($\beta_2=0.25$)+$^{92}$Zr($\beta_2=0.1$) [$Q_{2n}$=5.1 MeV]\\
and\\
$^{32}$S($\beta_2=0.31$)+$^{96}$Zr($\beta_2=0.08$)$\to ^{34}$S($\beta_2=0.25$)+$^{94}$Zr($\beta_2=0.09$) [$Q_{2n}$=5.7 MeV],\\
the deformation  of  S  slightly decreases  and
the values of the corresponding Coulomb barriers slightly increase.
As a result, the transfer
weakly  suppresses the capture process at the sub-barrier energies.
This suppression becomes  stronger with decreasing  energy.
One can see in Fig. 5 that
at energies above, near, and below the Coulomb barrier
the cross sections with and without two-neutron transfer are almost similar in the reactions $^{32}$S+$^{94,96}$Zr.
The relative enhancement of the sub-barrier fusion cross sections in
the reactions
$^{32}$S+$^{94,96}$Zr with respect to that in the reactions
$^{36}$S+$^{90,96}$Zr  is mainly related to the  deformation
of $^{34}$S in the $2^+$ state.
With respect to the reactions $^{36}$S+$^{94,96}$Zr
the enhancements of cross sections in the reactions $^{32}$S+$^{94,96}$Zr and $^{32}$S+$^{90}$Zr
are similar because of   the close  deformations of interacting nuclei after neutron transfer.
So, the observed capture enhancement at  sub-barrier energies in
the reactions $^{32}$S+$^{94,96}$Zr and $^{32}$S+$^{90}$Zr
 is not related to the  transfer effects but  to
the direct static deformation effects.


\section{Summary}
The quantum diffusion approach, the universal fusion function representation, the extracted
capture probabilities from the experimental excitation functions are applied to study
the role of the neutron transfer with   positive $Q_{xn}$-values
in the capture (fusion) reactions $^{40}$Ca+$^{94,96}$Zr and $^{32}$S+$^{94,96}$Zr.
We found that the change of   the
 capture (fusion) cross section after the two-neutron transfer
occurs due to the change of the deformations of  nuclei.
When after the neutron transfer the deformations of nuclei
strongly (weakly) change,
the neutron transfer  strongly (weakly) influences  the  fusion  cross section.
We clearly showed that the neutron transfer effects on the excitation functions
in the reactions $^{40}$Ca+$^{94,96}$Zr and $^{32}$S+$^{94,96}$Zr are completely different.
The calculations pointed  a strong increase of the fusion enhancement due to the neutron transfer
 for the systems with the spherical acceptor-nuclei as in the case of the reactions $^{40}$Ca+$^{94,96}$Zr.
 In the reactions $^{32}$S+$^{94,96}$Zr with the  well deformed acceptor-nucleus $^{32}$S,
 the strong fusion enhancement arises due to the static deformation effects.

Combining all our calculations within the quantum diffusion approach,
one can come to the following conclusions about the role of neutron transfer in the capture (fusion) reactions
with positive $Q_{xn}$-values for the neutron transfer.\\
(a) If the acceptor-nucleus is  spherical or slightly deformed (relatively strongly bound)
  nucleus and the donor-nucleus is the spherical or deformed nucleus,
the neutron transfer may lead to the strong capture (fusion) enhancement
[for example,  $^{40}$Ca+$^{94,96}$Zr,$^{116,124,132}$Sn,$^{154}$Sm,   $^{58}$Ni+$^{64}$Ni, and $^{40}$Ca+$^{48}$Ca]
or to the weak influence or even to the  weak suppression
[for example,  $^{18}$O+$^{92}$Mo,$^{112,118,124}$Sn].\\
(b) If the acceptor-nucleus is strongly deformed   (relatively weakly bound) nucleus
and the donor-nucleus is the spherical or deformed nucleus,
the neutron transfer may lead   to the weak influence or even to the  weak suppression
[for example, $^{18}$O+$^{74}$Ge, $^{60}$Ni+$^{100}$Mo, $^{28}$Si+$^{142}$Ce,$^{154}$Sm, and
$^{32}$S+$^{94,96}$Zr,$^{96-100}$Mo,$^{100-104}$Ru,$^{110}$Pd,$^{154}$Sm,$^{208}$Pb].
In these reactions with  strongly deformed nuclei $^{28}$Si, $^{32}$S, and $^{74}$Ge
the fusion enhancement is caused by the static deformation effects.

Thus, the  point of view that  the sub-barrier capture (fusion) cross section can be
  weakly influenced or even suppressed because of the neutron transfer
with    positive $Q_{xn}$-values has to be carefully  studied. We predict the weak
neutron transfer effects in the fusion reactions
$^{60,62}$Ni+$^{150}$Nd, $^{18}$O+$^{64}$Ni,$^{114,116,120,122,126}$Sn,$^{204,206}$Pb,
and $^{28}$Si,$^{32}$S+$^{116-134}$Sn,$^{148-152}$Sm.
As shown with the quantum diffusion approach,
 the capture cross sections almost match in the reactions
 $^{16}$O+$^{52}$Cr  and
 $^{18}$O+$^{50}$Cr,  
 $^{16}$O+$^{94}$Mo and
 $^{18}$O+ $^{92}$Mo, 
 $^{16}$O+$^{114,120,124,126}$Sn and
 $^{18}$O+$^{112,118,122,124}$Sn,
 respectively.
 The same reduced fusion cross sections for the reactions $^{58,60,62}$Ni+$^{150}$Nd with  positive
 $Q_{2n}$-values [$Q_{2n}$=8.2, 6.0, 4.1 MeV, respectively] are predicted.

G.G.A. and  N.V.A.   acknowledge  the partial
 support from the Alexander von Humboldt-Stiftung (Bonn).
This work was supported by   RFBR and NSFC.
The IN2P3(France)-JINR(Dubna)
Cooperation Programme  is gratefully acknowledged.\\


\begin{figure}
\includegraphics[scale=1]{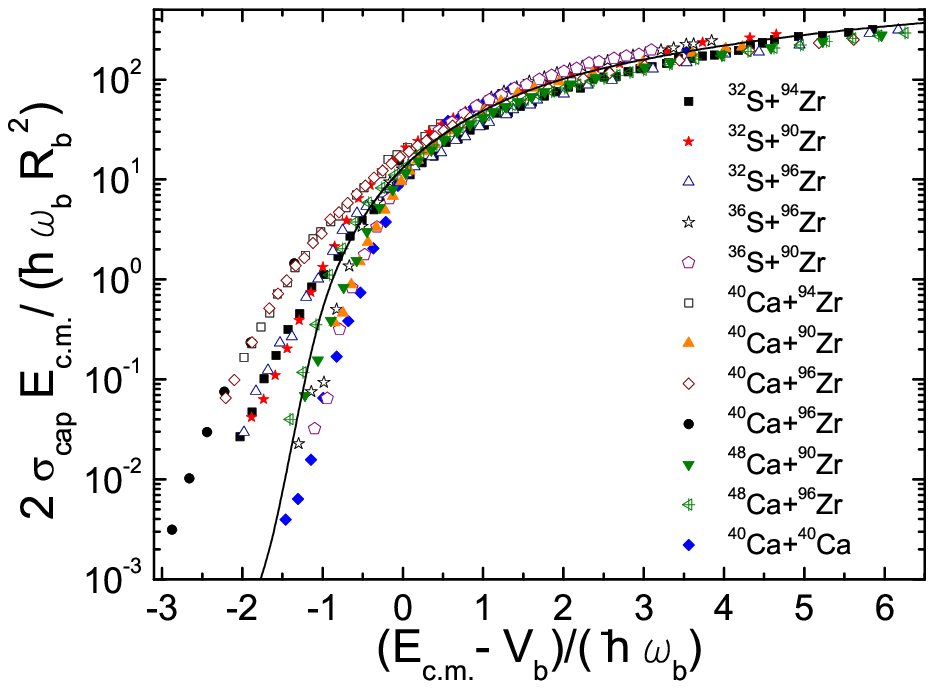}
\caption{(Color online) Comparison of the universal fusion function
(solid line) with the experimental
values~\protect\cite{TimmersCa40Zr96,StefaniniCa40Zn94,StefMont40ca96zr,ZhangS32Zr9096,Jia32s94zr,StefaniniCa48Zn9096,StefaniniS36Zn9096,Stefan40ca40ca}
(symbols) of $\dfrac{2E_{\rm c.m.}}{\hbar\omega_bR_b^2}\sigma_{cap}(E_{\rm c.m.})$
vs $\dfrac{E_{\rm c.m.}-V_b}{\hbar\omega_b}$ for the reactions  indicated.
}
\label{1 _fig}
\end{figure}

\begin{figure}
\includegraphics[scale=0.5]{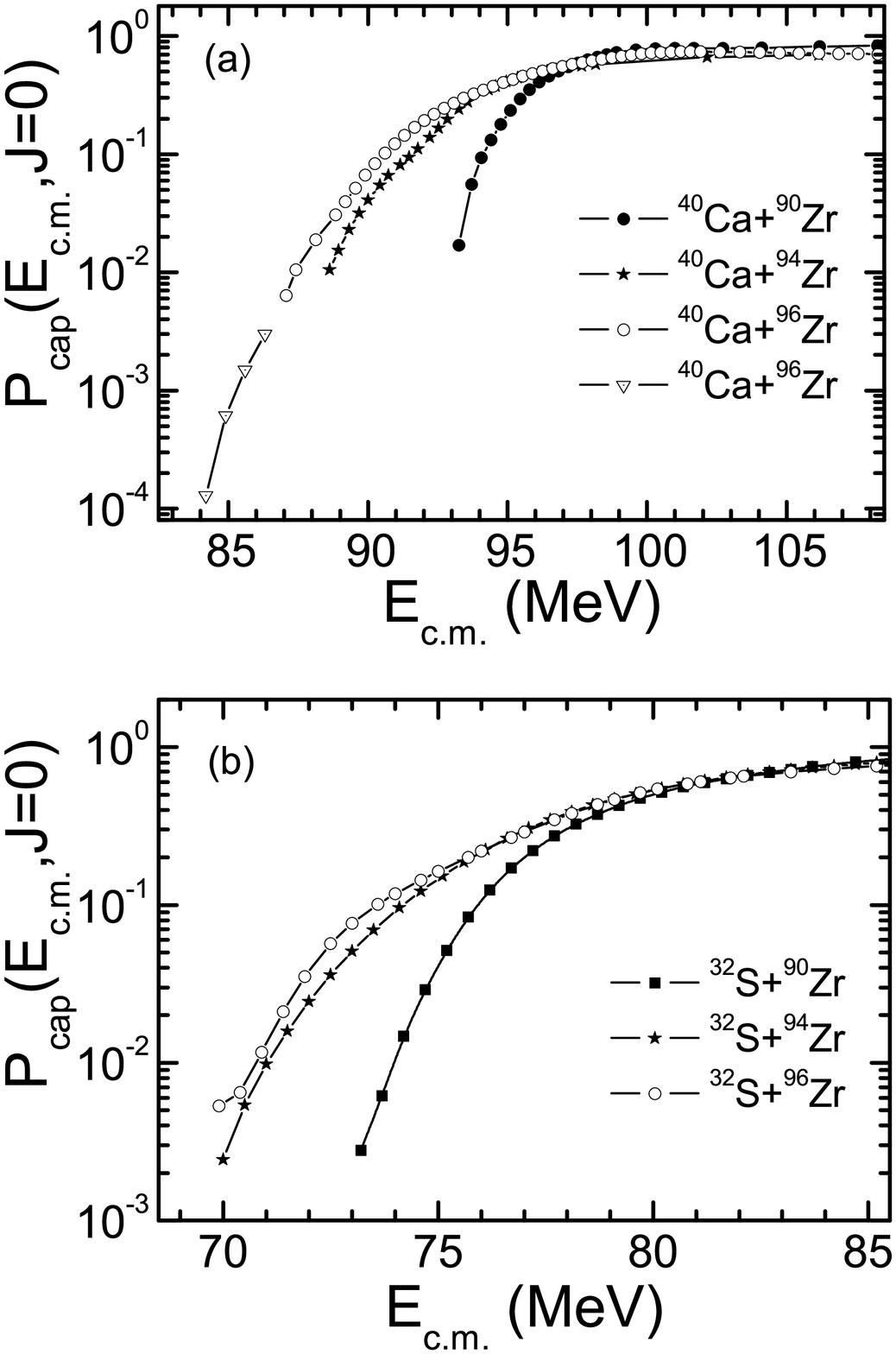}
\caption{
The extracted   $s$-wave  capture  probabilities  for the     reactions  indicated
by employing  Eq. (\ref{g3_eq})  [symbols connected by lines].
The used experimental   capture (fusion)
excitation functions are from Refs.~\protect\cite{TimmersCa40Zr96,StefaniniCa40Zn94,StefMont40ca96zr,ZhangS32Zr9096,Jia32s94zr}.
}
\label{2_fig}
\end{figure}

\begin{figure}
\includegraphics[scale=0.5]{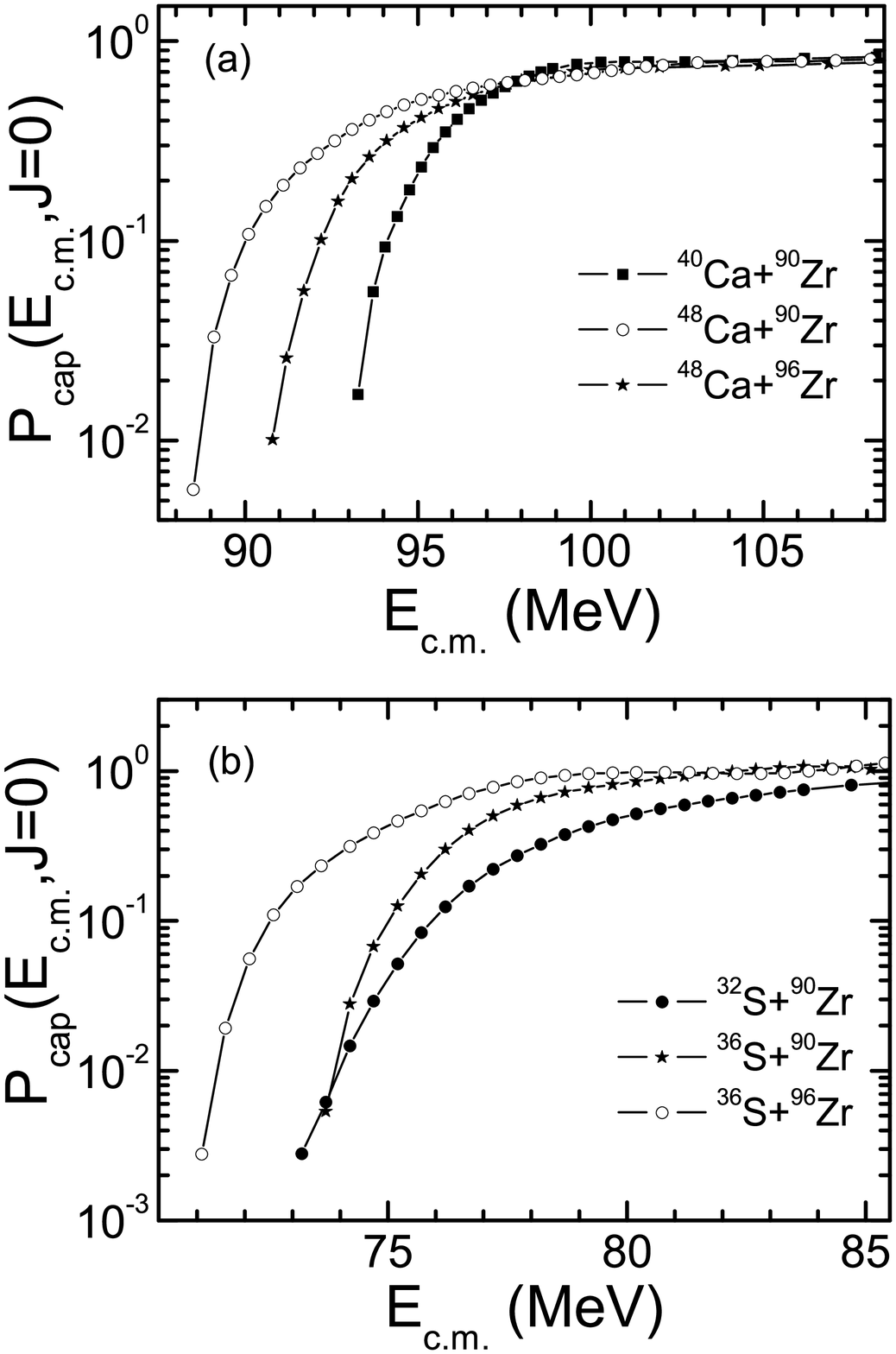}
\caption{
The extracted   $s$-wave  capture  probabilities  for the      reactions  indicated
by employing  Eq. (\ref{g3_eq})   [symbols connected by lines].
The used experimental   capture (fusion)
excitation functions are from
Refs.~\protect\cite{TimmersCa40Zr96,ZhangS32Zr9096,StefaniniCa48Zn9096,StefaniniS36Zn9096}.
}
\label{3_fig}
\end{figure}

\begin{figure}
\includegraphics[scale=0.5]{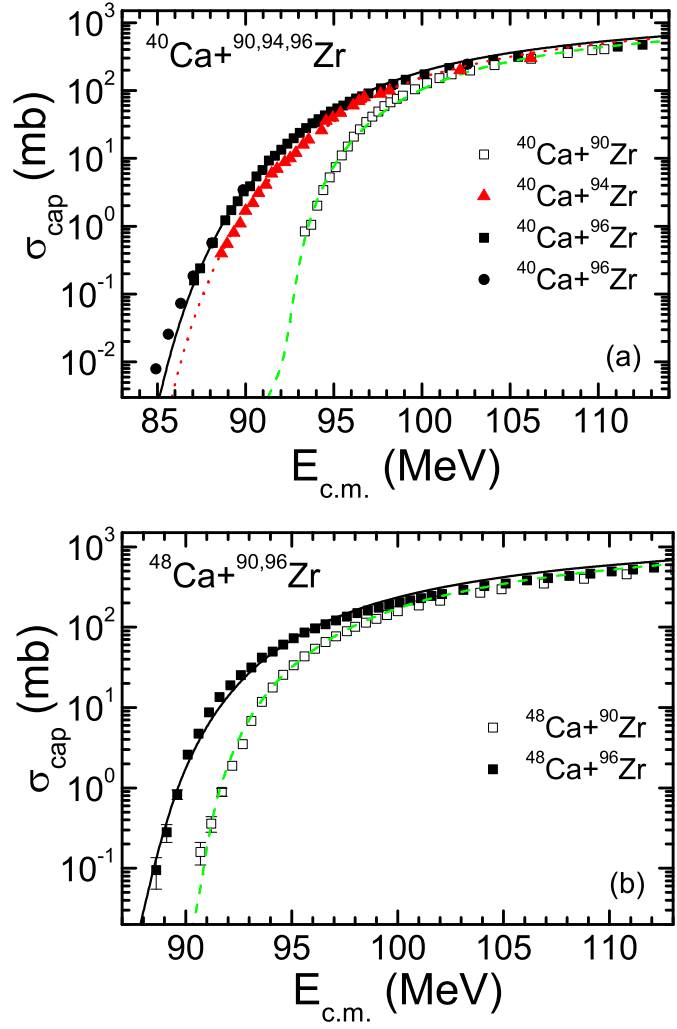}
\caption{(Color online)
The calculated capture cross sections vs $E_{\rm c.m.}$ for the reactions (a)
$^{40}$Ca+$^{96}$Zr (solid line), $^{40}$Ca+$^{94}$Zr (dotted line), $^{40}$Ca+$^{90}$Zr (dashed line)  and
(b) $^{48}$Ca+$^{96}$Zr (solid line), $^{48}$Ca+$^{90}$Zr (dashed line).
The experimental data (symbols) are from Refs.~\protect\cite{TimmersCa40Zr96,StefaniniCa40Zn94,StefMont40ca96zr,StefaniniCa48Zn9096}.
}
\label{4_fig}
\end{figure}

\begin{figure}
\includegraphics[scale=0.5]{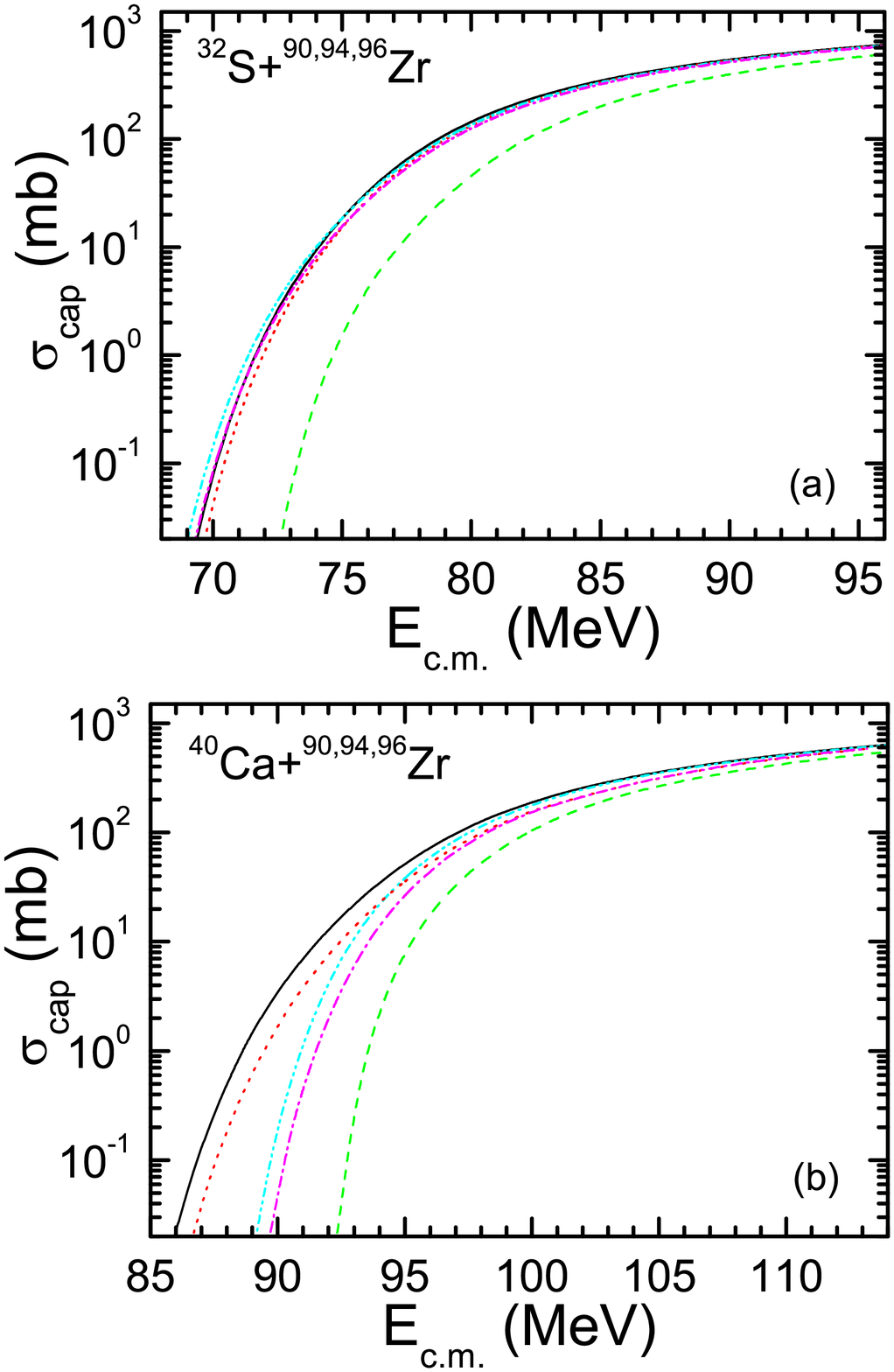}
\caption{(Color online)
The calculated capture cross sections vs $E_{\rm c.m.}$ for the  reactions (b)
$^{40}$Ca+$^{96}$Zr (solid line), $^{40}$Ca+$^{94}$Zr (dotted line), $^{40}$Ca+$^{90}$Zr (dashed line)  and (a)
$^{32}$S+$^{96}$Zr (solid line), $^{32}$S+$^{94}$Zr (dotted line), $^{32}$S+$^{90}$Zr (dashed line).
For the  reactions $^{32}$S,$^{40}$Ca+$^{96}$Zr  and $^{32}$S,$^{40}$Ca+$^{94}$Zr, the capture cross
sections calculated without  taking into consideration the neutron transfer
are shown by dash-dot-dotted and dash-dotted [it is matched with solid line in the part (a)] lines, respectively.
}
\label{5_fig}
\end{figure}

\begin{figure}
\includegraphics[scale=0.5]{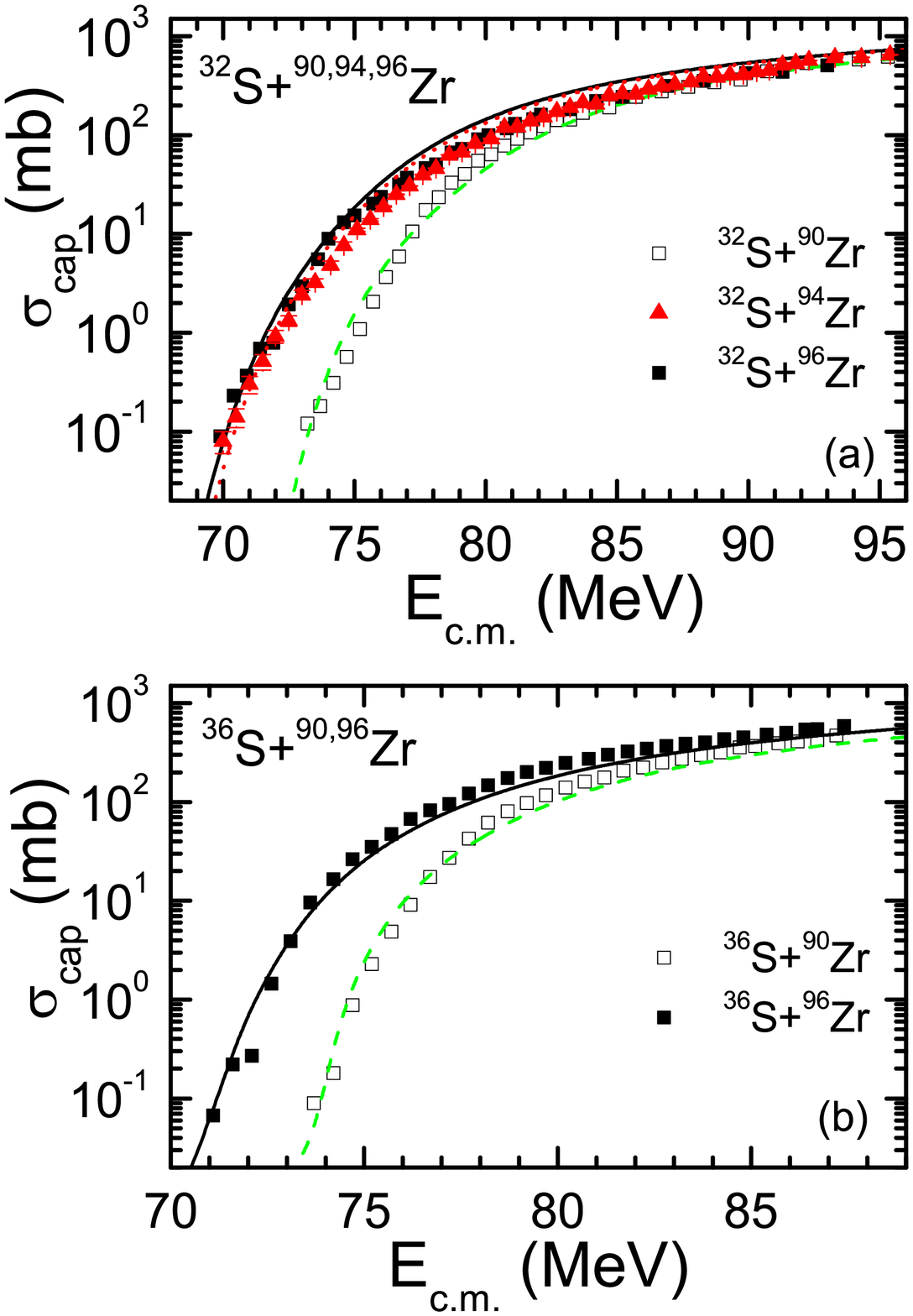}
\caption{(Color online)
The calculated capture cross sections vs $E_{\rm c.m.}$ for the reactions (a)
$^{32}$S+$^{96}$Zr (solid line), $^{32}$S+$^{94}$Zr (dotted line), $^{32}$S+$^{90}$Zr (dashed line)
and
(b) $^{36}$S+$^{96}$Zr (solid line), $^{36}$S+$^{90}$Zr (dashed line).
The experimental data (symbols) are from Refs.~\protect\cite{ZhangS32Zr9096,Jia32s94zr,StefaniniS36Zn9096}.
}
\label{6_fig}
\end{figure}


\begin{thebibliography}{99}

\bibitem{Gomes} L.F.~Canto, P.R.S.~Gomes, R.~Donangelo, and M.S.~Hussein, Phys. Rep. {\bf 424}, (2006) 1  and references  therein.

\bibitem{Back} B.B. Back, H. Esbensen, C.L. Jiang and K.E. Rehm, Rev. Mod. Phys. \textbf{86}, 317(2014) and references  therein.

\bibitem{Beckerman} M.~Beckerman {\it et al.}, Phys. Rev. Lett. {\bf 45} (1980) 1472;
M.~Beckerman, J.~Ball, H.~Enge,  M.~Salomaa, A.~Sperduto, S.~Gazes, A.~DiRienzo, and J.D.~Molitoris,
Phys. Rev. C {\bf 23} (1981) 1581;
M.~Beckerman, M.~Salomaa, A.~Sperduto, J.D.~Molitoris, and A.~DiRienzo,
Phys. Rev. C {\bf 25} (1982) 837.
\bibitem{Broglia} R.A.~Broglia, C.H.~Dasso, S.~Landowne, and A.~Winther,
Phys. Rev. C {\bf 27}, 2433 (1983);
R.A.~Broglia, C.H.~Dasso, S.~Landowne, and  G.~Pollarolo,
 Phys. Lett. B  {\bf 133}, 34 (1983).

\bibitem{TimmersCa40Zr96}    H.~Timmers {\it et al.}, Nucl. Phys.  {\bf A633}, 421 (1998).
\bibitem{StefaniniCa40Zn94}     A.M.~Stefanini~{\it et al.},  Phys. Rev. C {\bf 76}, 014610 (2007).
\bibitem{StefMont40ca96zr} A.M.~Stefanini~{\it et al.},     Phys. Lett. B {\bf 728}, 639 (2014).

\bibitem{Stefanini40ca116124sn}
F.~Scarlassara {\it et al.}, Nucl. Phys.  {\bf A672}, 99 (2000).

\bibitem{Kol40Ca124132Sn} J.J.~Kolata {\it et al.}, Phys. Rev. C {\bf 85}, 054603 (2012).



\bibitem{Pengo} R.~Pengo {\it et al.}, Nucl. Phys. {\bf A411}, 255 (1983).

\bibitem{Henning} W.~Henning, F.L.H.~Wolfs, J.P.~Schiffer, and K.E.~Rehm,
Phys. Rev. Lett. {\bf 58}, 318 (1987).
\bibitem{Stelson} P.H.~Stelson, H.J.~Kim, M.~Beckerman, D.~Shapira, and R.L.~Robinson,
Phys. Rev. C {\bf 41}, 1584 (1990).
\bibitem{Roberts}  R.B.~Roberts {\it et al.}, Phys.
Rev. C {\bf 47}, R1831 (1993).
\bibitem{Stefanini3236s110pd} A.M.~Stefanini~{\it et al.},
Phys. Rev. C {\bf 52}, R1727 (1995).
\bibitem{Sonzogni} A.A.~Sonzogni,
 J.D.~Bierman, M.P.~Kelly, J.P.~Lestone, J.F.~Liang, and R.~Vandenbosch, Phys.
Rev. C {\bf 57}, 722 (1998).
\bibitem{Tripathi} V.~Tripathi  {\it et al.}, Phys. Rev. C {\bf 65}, 014614  (2001).

\bibitem{Jiang40ca48ca}        C.L.~Jiang~{\it et al.}, Phys. Rev. C {\bf 82},  041601(R) (2010).

\bibitem{ZhangS32Zr9096} H.Q.~Zhang~{\it et al.}, Phys. Rev. C {\bf 82}, 054609 (2010).

\bibitem{Kola40ca134Te} Z.~Kohley {\it et al.}, Phys. Rev. C {\bf 87}, 064612 (2013).

\bibitem{Jia32s94zr} H. M. Jia  {\it et al.}, Phys. Rev. C {\bf 89}, 064605 (2014).


\bibitem{AOASn} P. Jacobs, Z. Fraenkel, G. Mamane, and L. Tserruya, Phys.  Lett. B {\bf 175},  271 (1986).
\bibitem{Scarlassara} F.~Scarlassara {\it et al.}, EPJ Web Conf. {\bf 17}, 05002 (2011).
\bibitem{Liang}      Z.~Kohley {\it et al.}, Phys. Rev. Lett. {\bf 107},  202701 (2011);
J.F.~Liang, EPJ Web Conf. {\bf 17}, 02002 (2011);
J.F.~Liang {\it et al.}, Phys. Rev. C {\bf 78}, 047601 (2008).

\bibitem{Jia18O74Ge} H.M.~Jia {\it et al.}, Phys. Rev. C {\bf 86}, 044621 (2012).

\bibitem{Jiang2014} C.L. Jiang  {\it et al.}, Phys. Rev. C {\bf 89}, 051603(R) (2014).

\bibitem{EPJSub}  V.V.~Sargsyan, G.G. Adamian, N.V. Antonenko, and W. Scheid,
Eur. Phys. J. A {\bf 45}, 125 (2010);
V.V.~Sargsyan, G.G.~Adamian, N.V.~Antonenko,  W.~Scheid, and H.Q.~Zhang,
Eur. Phys. J. A {\bf 47}, 38 (2011); J. of Phys.: Conf. Ser. {\bf 282}, 012001 (2011);
EPJ Web Conf. {\bf 17}, 04003 (2011);
V.V.~Sargsyan, G.G.~Adamian, N.V.~Antonenko,  W.~Scheid, C.J.~Lin,
and H.Q.~Zhang, Phys. Rev. C {\bf 85}, 017603 (2012);
Phys. Rev. C {\bf 85}, 037602 (2012).
\bibitem{EPJSub1}        V.V.~Sargsyan, G.G.~Adamian, N.V.~Antonenko,
W.~Scheid, and H.Q.~Zhang, Phys. Rev. C {\bf 84}, 064614 (2011);
Phys. Rev. C {\bf 85}, 024616 (2012).

\bibitem{EPJSub3} V.V.~Sargsyan, G.G.~Adamian, N.V.~Antonenko,  W.~Scheid, and H.Q.~Zhang,
{\it   Phys. Rev.}  C {\bf 86},  014602 (2012).

\bibitem{EPJSub21}   V.V.~Sargsyan, G.G.~Adamian, N.V.~Antonenko,  W.~Scheid, and H.Q.~Zhang,
{\it  Eur. Phys. J.} A  {\bf 49},  54 (2013).

\bibitem{SD} V.V.~Sargsyan, G.~Scamps, G.G.~Adamian, N.V.~Antonenko,  and D.~Lacroix,
Phys. Rev. C {\bf 88},  064601  (2013).


\bibitem{Gomes1} L.F.~Canto, P.R.S.~Gomes, J.~Lubian, L.C.~Chamon, and E.~Crema,
                        J. Phys. G {\bf 36}, 015109 (2009).
\bibitem{Gomes2} L.F.~Canto, P.R.S.~Gomes, J.~Lubian, L.C.~Chamon, and E.~Crema,
                        Nucl. Phys.  {\bf A821}, 51 (2009).

\bibitem{StefaniniCa48Zn9096}     A.M.~Stefanini~{\it et al.},  Phys. Rev. C {\bf 73}, 034606 (2006).

\bibitem{StefaniniS36Zn9096}   A.M.~Stefanini~{\it et al.},  Phys. Rev. C {\bf 62}, 014601 (2000).

\bibitem{Stefan40ca40ca} G.~Montagnoli,~{\it et al.}, Phys. Rev. C {\bf 85},  024607 (2012).

\bibitem{Bala} A.B.~Balantekin, S.E.~Koonin, and J.W.~Negele, Phys. Rev. C
\textbf{28}, 1565 (1983); A.B.~Balantekin  and P.E.~Reimer, Phys. Rev. C
\textbf{33}, 379 (1986);  A.B.~Balantekin, A.J.~DeWeerd, and S.~Kuyucak, Phys. Rev. C
\textbf{54}, 1853 (1996).

\bibitem{Bezier}        P. Henrici, {\it Essentials of numerical analysis}, (John Wiley \& Sons, Inc. New York, NY, USA, 1982).

\bibitem{PRCPOP} R.A.~Kuzyakin, V.V.~Sargsyan, G.G.~Adamian, N.V.~Antonenko, E.E.~Saperstein, and S.V.~Tolokonnikov,
Phys. Rev. C {\bf 85}, 034612 (2012).


%
%
\bibitem{SSzilner} S.~Szilner {\it et al.}, Phys. Rev. C {\bf 76}, 024604 (2007); S.~Szilner {\it et al.}, Phys. Rev. C {\bf 84}, 014325 (2011).
%
%
%
\bibitem{Ram}           S.~Raman, C.W.~Nestor, Jr, and P.~Tikkanen, At. Data Nucl. Data Tables {\bf 78}, 1 (2001).




\end{thebibliography}
\end{document}